\documentclass{PoS}

\title{QCD thermodynamics with O(a) improved Wilson fermions at $N_f=2$}

\ShortTitle{QCD thermodynamics with O(a) improved Wilson fermions at $N_f=2$}

\author{\speaker{Bastian B. Brandt} \\
        Institut f\"ur theoretische Physik, \\
        Universit\"at Regensburg,
        D-93040 Regensburg\\
        E-mail: \email{bastian.brandt@physik.uni-regensburg.de}}

\author{Anthony Francis, Harvey B. Meyer and Hartmut Wittig \\
        PRISMA Cluster of Excellence, Institut f\"ur Kernphysik \\
        and Helmholtz Institut Mainz, \\
        Johannes Gutenberg-Universit\"at Mainz,
        D-55099 Mainz}

\author{Owe Philipsen\\
        Institut f\"ur Theoretische Physik, \\
        Goethe-Universit\"at,
        D-60438 Frankfurt am Main}

\abstract{We present an update of our study of the phase diagram of two-flavour
QCD at zero baryon density with dynamical $O(a)$ improved Wilson quarks. All
simulations are done on lattices with a temporal extent of $N_t=16$ and spatial
extent $L=32,48$ and 64, ensuring that discretisation effects are small and
finite size effects can be controlled. In the approach to the chiral limit we
currently have three scans with pion masses between 540 and 200~MeV. In this
proceedings article the focus is on the new scan at $m_\pi=200$~MeV and the
measurement of screening masses. We also present first results concerning a
test of scaling in the approach to the chiral limit and the chiral extrapolation
of the difference of screening masses in scalar and pseudoscalar channels, which
provides a measure for the strength of the anomalous breaking of the $U_A(1)$
symmetry.}

\FullConference{31st International Symposium on Lattice Field Theory LATTICE
2013\\
         July 29 -- August 3, 2013\\
         Mainz, Germany}

\begin{document}

\section{Introduction}

Due to the continued progress in the field of lattice QCD at finite temperature
in the last few years, most of the features of the phase diagram in the
$\{m_{ud},m_s,T\}$ parameter space at vanishing chemical potential are by now
rather well understood (for a review see~\cite{Lombardo:2012ix}). One of the
most prominent remaining issues at vanishing chemical potential concerns
the behaviour of the chiral critical line with increasing strange quark mass.
There are two possible scenarios~\cite{Pisarski:1983ms,Butti:2003nu}: Either the
chiral critical line reaches the $m_{ud}=0$ axis at some tri-critical point, so
that from this point on the transition is of second-order in the chiral limit,
or the chiral critical line continues at non-vanishing $m_{ud}$ up to infinitely
heavy strange quarks, leaving a first-order transition at $m_{ud}=0$. The
universality class of the chiral transition in the second-order scenario depends
on the strength of the anomalous breaking of the $U_A(1)$
symmetry~\cite{Pisarski:1983ms,Butti:2003nu,Pelissetto:2013hqa}. In
case of a strong breaking the transition is in the 3d $O(4)$ universality
class~\cite{Pelissetto:2013hqa,Rajagopal:1992qz}, for a weak breaking in the 3d
$U(2)\times U(2) \to U(2)$ universality
class~\cite{Butti:2003nu,Pelissetto:2013hqa}. The first-order scenario is
expected to be realised for a nearly restored $U_A(1)$ symmetry.

Here we present an update of our
study~\cite{Brandt:2010uw,Brandt:2010bn,Brandt:2012sk} concerning the order of
the transition in the chiral limit of the two-flavour theory with
non-perturbatively $O(a)$ improved Wilson fermions. To investigate the approach
to the chiral limit with control over the main systematic effects we use
$N_t=16$ throughout, thereby reducing the cutoff effects to a minimum, and aim
at simulations at several pion masses below
300~MeV at three different volumes each. In this proceedings article the focus
will be on the measurements of screening masses and first attempts to
extrapolate our results to the chiral limit. For details concerning
strategy and setup we refer to our earlier publications. Another aspect of our
project is the measurement of plasma properties in the transition region and
above~\cite{Brandt:2012jc,Brandt:2013fg,Brandt:2013faa}, which will not be
covered here.

\section{Temperature scans and transition temperatures}

\subsection{Scan setup}

Our simulations use non-perturbatively $O(a)$ improved Wilson
fermions~\cite{Sheikholeslami:1985ij} with two degenerate dynamical
quarks and the Wilson plaquette action. The configurations are generated with
deflation accelerated DD-HMC~\cite{Luscher:2005rx,Luscher:2007es} and
MP-HMC~\cite{Marinkovic:2010eg} algorithms. The lattice sizes are chosen to be
$16\times32^3$, $48^3$ and $64^3$ to suppress cutoff effects and to enable the
extrapolation to the thermodynamic limit. To scan in temperature we vary the
bare coupling $\beta$ either at fixed hopping parameter $\kappa$ (for
$m_\pi>300$~MeV) or fixed renormalised quark mass (for $m_\pi<300$~MeV). Scale
setting, renormalisation and tuning for lines of constant physics is done using
input from CLS~\cite{Leder:2010kz,Fritzsch:2012wq,Brandt:2013dua}. To determine
the mass scale we use the renormalised PCAC mass in the
$\overline{\textnormal{MS}}$ scheme at $\mu=2$~GeV measured directly on our
finite-T ensembles (see~\cite{Brandt:2013faa}).

Our main observables concerning
the transition are the real part of the APE-smeared Polyakov loop ($L_{\rm
SM}$), the subtracted chiral condensate~\cite{Bochicchio:1985xa,Giusti:1998wy}
($\left<\bar{\psi}\psi\right>_{\rm sub}$) as defined in~\cite{Brandt:2013fg}
and the associated susceptibilities. At the moment those observables are
unrenormalised. The error analysis has been done using the bootstrap method with
1000 bins.

\subsection{Transition temperatures}

\begin{table}[t]
\vspace*{-3mm}
\begin{center}
\small
\begin{tabular}{c|ccccc|cc}
\hline
\hline
scan & Lattice & $\kappa$/$m_{ud}$~[MeV] & $T$~[MeV] &
$\tau_{U_P}$~[MDU] & MDUs & $T_C$~[MeV] & $(m_{ud})_C$~[MeV] \\
\hline
{\bf B1}$_\kappa$ & $16\times32^3$ & 0.136500 & $190-275$ &
$\sim10$ & $\sim20000$ & 245 (7)(6) & 45 ( 2)( 2) \\
{\bf B3}$_\kappa$ & $16\times64^3$ & 0.136500 & $240-250$ &
$\sim46$ & $\sim16000$ & & \\
\hline
{\bf C1} & $16\times32^3$ & $\sim14.5$ & $150-250$ & $\sim8$ &
$\sim12000$ & 211 (5)(3) & 14.3 (14)( 9) \\
\hline
{\bf D1} & $16\times32^3$ & $\sim7.5$ & $175-250$ & $\sim8$ &
$\sim10000$ & 193 (7)(5) & 7.6 (19)( 7) \\
\hline
\hline
\end{tabular}
\end{center}
\vspace*{-3mm}
\caption{Scans at $N_t=16$ at constant $\kappa$ (subscript $\kappa$) and
constant renormalised quark mass. Listed is the temperature range in MeV, the
integrated autocorrelation time of the plaquette $\tau_{U_P}$, the number of
molecular dynamics units (MDUs) used for the analysis (the measurement frequency
being 4 MDUs) and the parameters, temperature $T_C$ and quark mass $(m_{ud})_C$,
at the critical point. For the latter, the first uncertainty is obtained from
the spread of results in the transition region. The second error contains the
uncertainties of the individual measurements, scale setting and
renormalisation.}
\label{tab:sims}
\end{table}

\begin{figure}[t]
\vspace*{-3mm}
 \centering
\includegraphics[width=.7\textwidth]{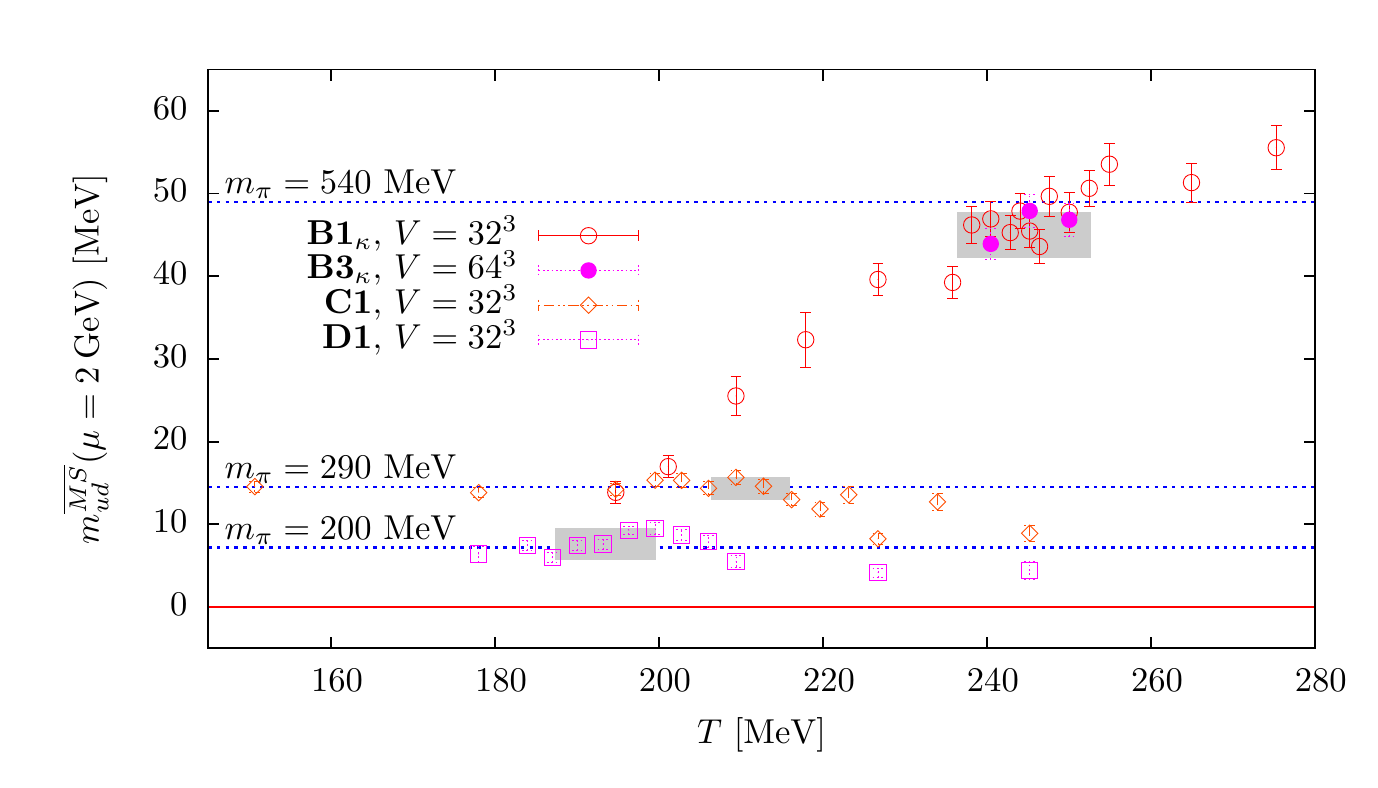}
\vspace*{-3mm}
 \caption{Simulation points in the $\{m_{ud},T\}$ parameter space. The grey
areas mark the crossover regions.}
 \label{fig:sims}
\end{figure}

\begin{figure}[t]
\vspace*{-5mm}
\begin{minipage}[c]{.48\textwidth}
\centering
\includegraphics[width=.95\textwidth]{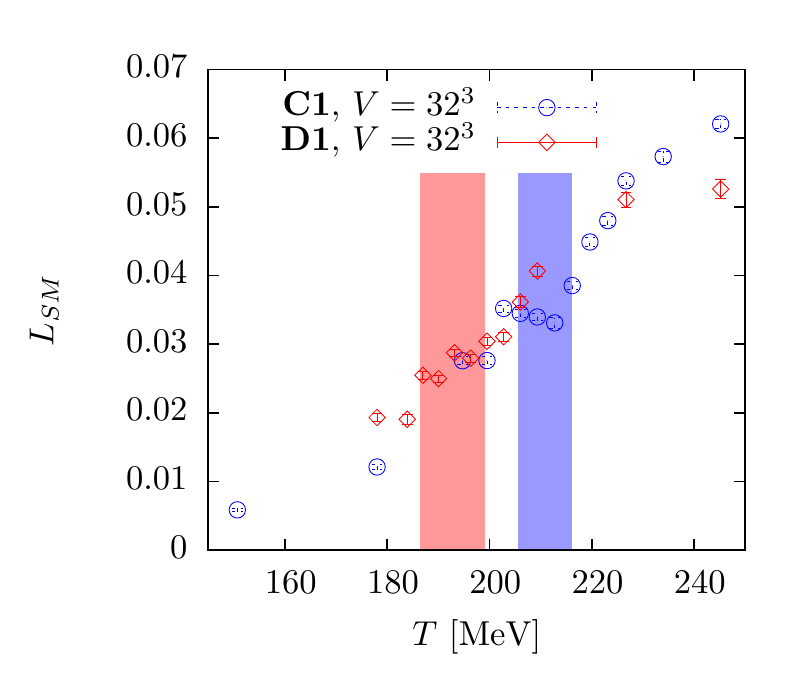}
\end{minipage}
\begin{minipage}[c]{.48\textwidth}
\centering
\includegraphics[width=.95\textwidth]{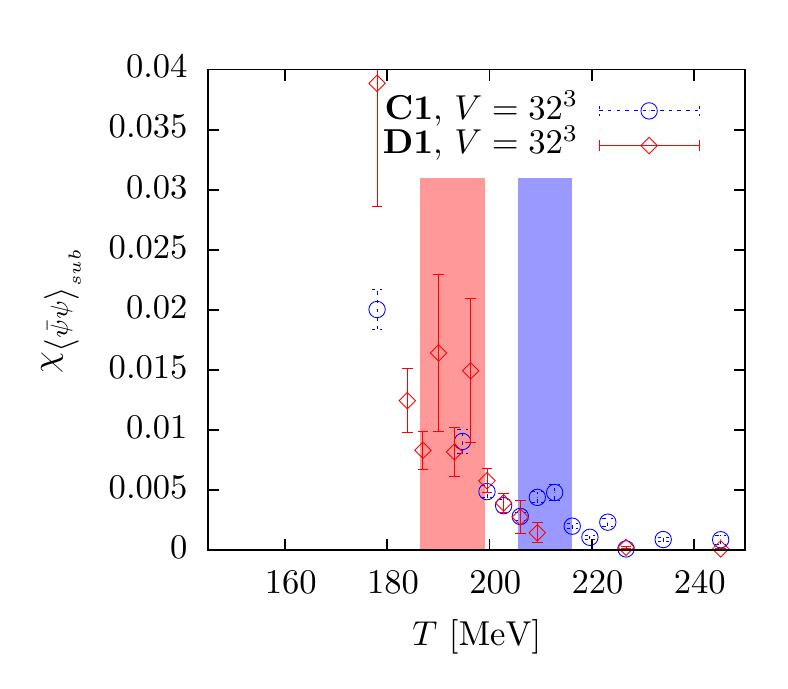}
\vspace*{-4mm}
\end{minipage}
\caption{Results for the smeared Polyakov Loop (left) and the susceptibility of
the subtracted chiral condensate (right) from scans {\bf C1} and {\bf D1}. The
filled areas mark the estimates for the transition regions.}
\label{fig:scan}
\end{figure}

So far our set of temperature scans at $N_t=16$ consists of three scans with
the simulation parameters given in table~\ref{tab:sims} and the associated
simulation points in the $\{m_{ud},T\}$ parameter space are displayed in
figure~\ref{fig:sims}. Scans {\bf B1}$_{\kappa}$, {\bf B3}$_\kappa$ and
{\bf C1} have already been presented in our earlier
publications~\cite{Brandt:2010uw,Brandt:2010bn,Brandt:2012sk} and the resulting
transition temperatures are given in table~\ref{tab:sims}, too~\footnote{The
results for scan {\bf C1} in table~\ref{tab:sims} and figure~\ref{fig:scan} have
been obtained with enlarged temperature range and statistics in comparison to
our earlier results presented in~\cite{Brandt:2012sk} and differ slightly.}. The
scan {\bf D1} is new and we show the results for the Polyakov loop and the
susceptibility of the subtracted chiral condensate in figure~\ref{fig:scan} in
comparison to the results from scan {\bf C1}.

The qualitative behaviour of the Polyakov loop is similar in both scans with a
slight shift towards smaller temperature values for scan {\bf D1}, indicating a
smaller transition temperature, in agreement with the expectations for smaller
quark masses. The susceptibilities of the subtracted condensate in the two scans
show a decrease when going to higher temperatures with a peak on top, due to the
superposition of the general $T=0$ behaviour (which is not subtracted here) and
the thermal effects. The transition temperature is extracted from the peak
using a fit to a Gaussian (see~\cite{Brandt:2012sk}). For scan {\bf D1} the
susceptibility still shows rather large fluctuations in the estimated transition
region and we currently increase statistics in this region to obtain a better
signal for $\chi_{\left<\bar{\psi}\psi\right>_{\rm sub}}$. Nevertheless, all
observables unambiguously indicate critical behaviour in the region between 186
and 200~MeV, which we identify as the transition region in table~\ref{tab:sims}.

\subsection{Scaling of the critical temperature}

With three transition temperatures at our disposal we can perform a first
scaling analysis in the approach to the chiral limit. In the continuum scaling
(which is the one realised for Wilson fermions due to explicit chiral symmetry
breaking at finite lattice spacing) the transition temperature changes with the
renormalised quark mass according to~\cite{Karsch:1993tv,Iwasaki:1996ya}
\begin{equation}
\label{eq:temp-scaling}
 T_C(m_{ud}) = T_C(0) \; \left[ 1 + C \; m_{ud}^{1/(\delta\beta)} \right] \;,
\end{equation}
Here $\delta$ and $\beta$ are the critical exponents associated with the
universality class of the critical point. This scaling will be realised up to
scaling violations in the approach to a critical point at $m_{ud}=0$, i.e. in
either of the second-order scenarios depicted in the introduction. The situation
is more complicated in the case of a first-order transition in the chiral limit,
due to the absence of an order parameter at the $Z(2)$ critical point at finite
$m_{ud}=m_{ud}^C$ (for a recent discussion see~\cite{Burger:2011zc}). Assuming
that eq.~(\ref{eq:temp-scaling}) also holds in the approach to this critical
point, one can simply replace $m_{ud}\to m_{ud}-m_{ud}^C$ in the associated
scaling analysis.

\begin{figure}[t]
\vspace*{-1mm}
\begin{minipage}[c]{.33\textwidth}
\centering
\includegraphics[width=.95\textwidth]{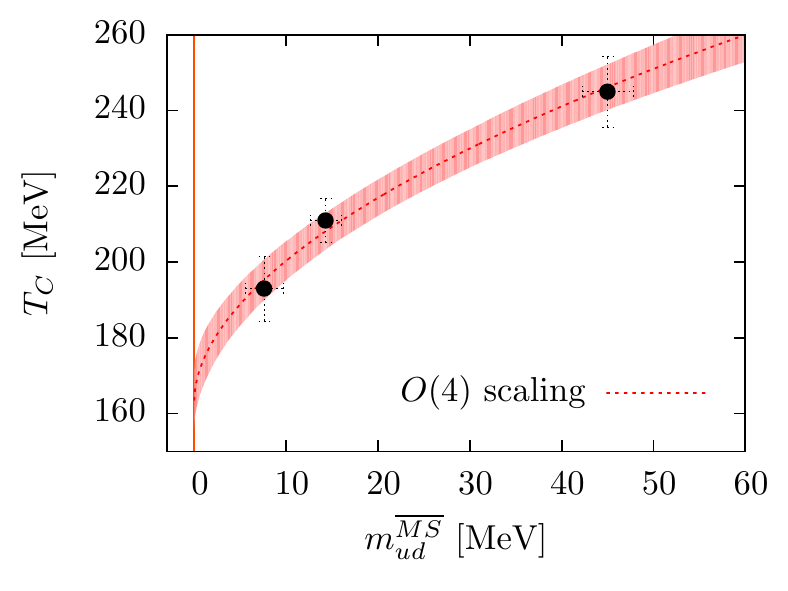}
\end{minipage}
\begin{minipage}[c]{.33\textwidth}
\centering
\includegraphics[width=.95\textwidth]{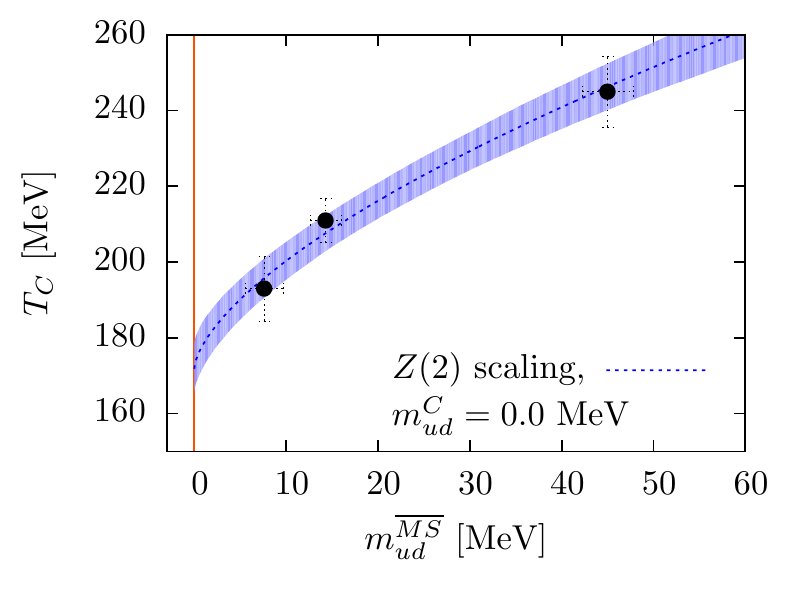}
\end{minipage}
\begin{minipage}[c]{.33\textwidth}
\centering
\includegraphics[width=.95\textwidth]{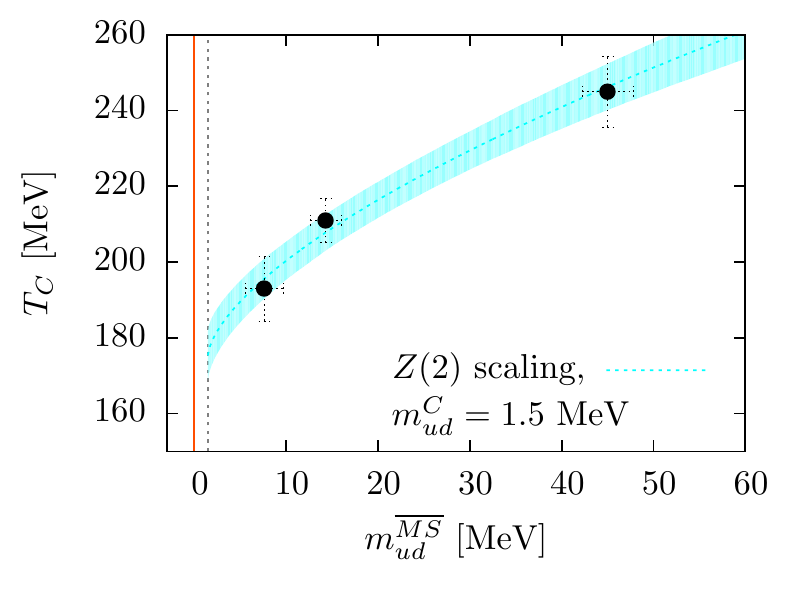}
\end{minipage}
\vspace*{-3mm}
\caption{Results for the scaling of the transition temperature using the
critical exponents of the $O(4)$ (left) and the $Z(2)$ universality class
(middle and right) at different values of $m_{ud}^C$.}
\label{fig:scaling}
\end{figure}

The precision of our data does not allow to constrain either the critical
exponents and/or $m_{ud}^C$, so that we are left with comparing the agreement
with the data using either of the critical exponents and different cases for
$m_{ud}^C$ (here $m_{ud}^C$=0 and 1.5~MeV) in the first-order scenario. The
results of the fits with fit parameters $T_C(0)$ and $C$ are shown in
figure~\ref{fig:scaling}. Since the critical exponents of the
$O(4)$ and the $U(2)\times U(2) \to U(2)$ scenario~\cite{Pelissetto:2013hqa} are
too similar to be distinguished, we have only shown the analysis for the $O(4)$
case. The plots demonstrate that all scenarios are compatible with the data.
This is not surprising given the still relatively large quark masses used for
the analysis. Similar problems to distinguish the scenarios from scaling alone
have been reported by the tmfT collaboration with pion masses between 300 and
500~MeV~\cite{Burger:2011zc}.

\section{Screening masses and $U_A(1)$ symmetry breaking}

To investigate the chiral symmetry restoration pattern we look at the degeneracy
of iso-vector screening masses in pseudoscalar ($P$), scalar ($S$), vector ($V$)
and axial vector ($A$) channels, the former two being directly related to the
breaking of the $U_A(1)$ symmetry~\cite{DeTar:1987ar}. The associated
correlation functions have been measured each 40 and 20 MDUs, for scans {\bf C1}
and {\bf D1}, respectively, using 15 point sources at different positions for
each configuration.

\begin{figure}[t]
\vspace*{-3mm}
\begin{minipage}[c]{.48\textwidth}
\centering
\includegraphics[width=.95\textwidth]{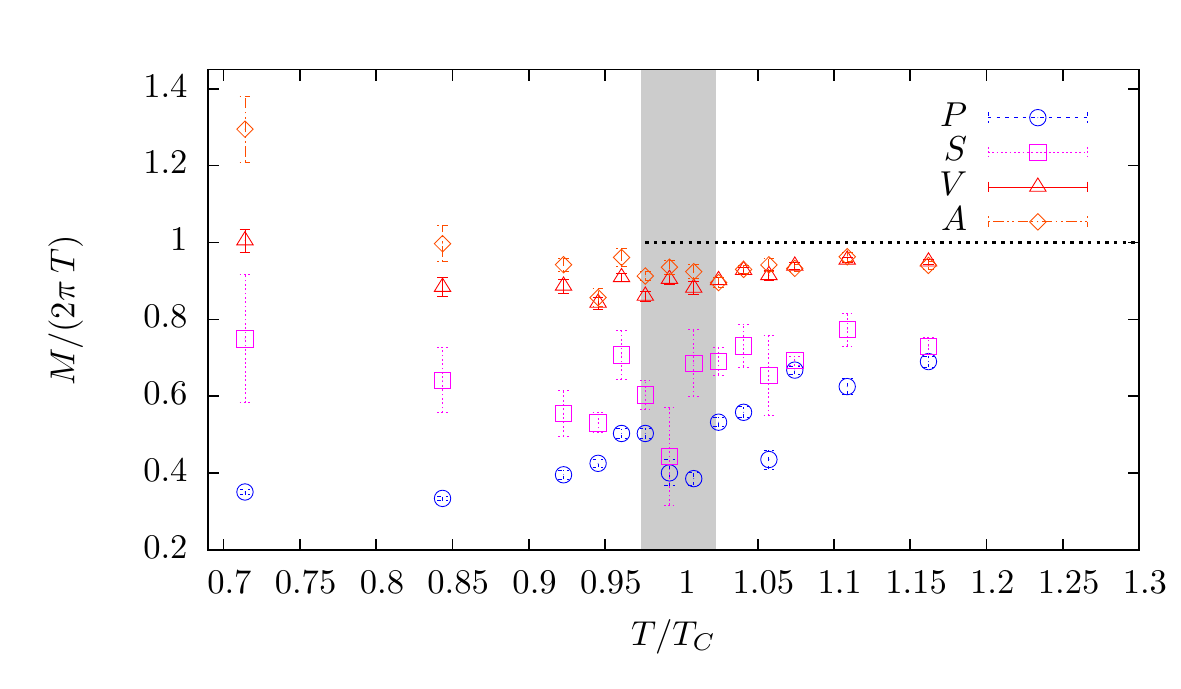} \\[-3mm]
\includegraphics[width=.95\textwidth]{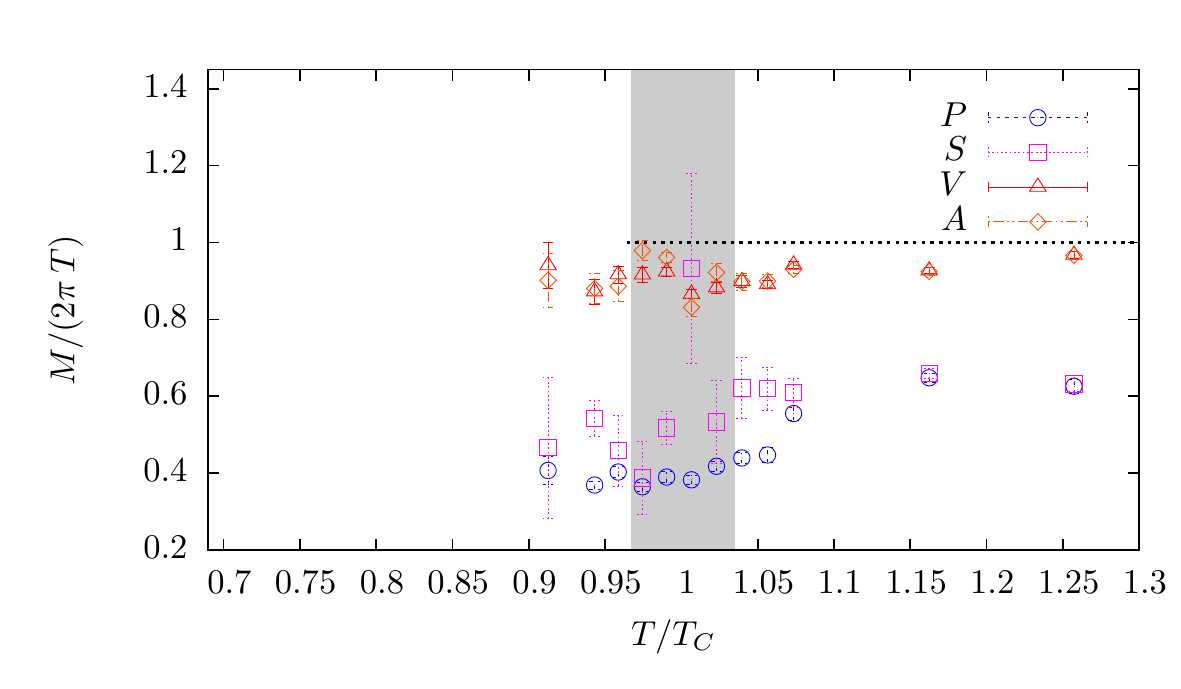}
\end{minipage}
\begin{minipage}[c]{.48\textwidth}
\centering
\includegraphics[width=.95\textwidth]{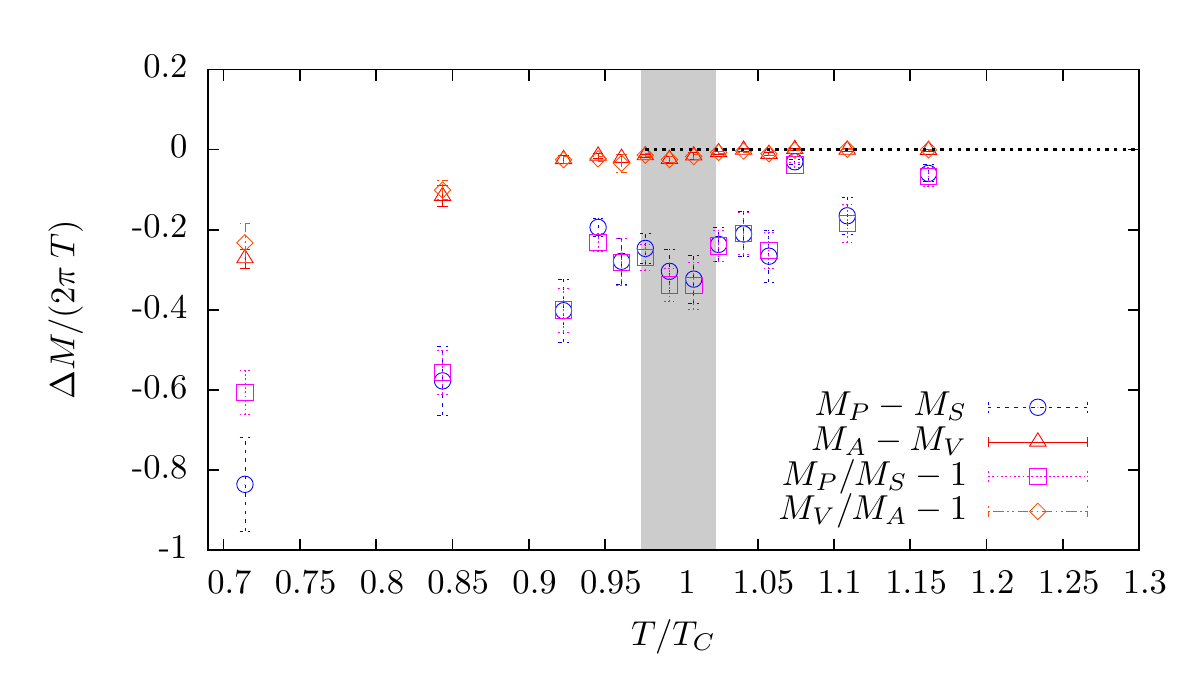} \\[-3mm]
\includegraphics[width=.95\textwidth]{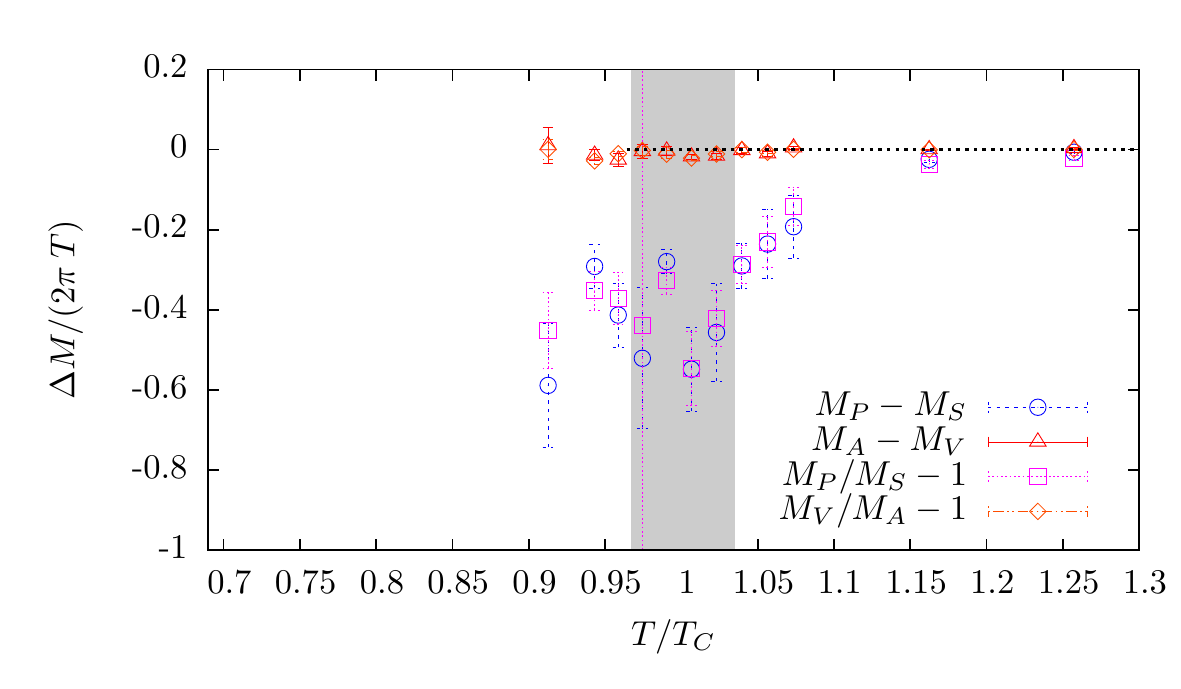}
\end{minipage}
\vspace*{-3mm}
\caption{{\bf Left:} Temperature dependence of screening masses in $P$, $S$,
$V$ and $A$ channels for scans {\bf C1} (top) and {\bf D1} (bottom). The dashed
line indicates the asymptotic value of $M=2\pi\:T$ for $T\to\infty$. {\bf
Right:} Screening mass differences in $S$ and $P$ and $A$ and $V$ channels in
units of $2\pi T$. The two sets of points have been extracted from fits to the
$x^3$-dependence of the mass difference and the mass ratio, respectively.}
\label{fig:screening}
\end{figure}

The temperature dependence of the screening masses is shown in
figure~\ref{fig:screening} (left) in comparison with the respective mass
splittings (right)~\footnote{The splittings are obtained from a fit to either
the effective mass for the difference $\Delta M$ or a fit to the ratio of
effective masses directly. This provides a more solid estimate for the mass
splittings since the plateau for these quantities are reached earlier and parts
of the fluctuations cancel.}. The screening masses start from the expected
splittings from the meson spectrum at zero-temperature (at least for scan {\bf
C1} where results at lower temperatures are available). In the approach to $T_C$
the screening masses in $S$ and $A$ channels become smaller while the ones in
$P$ and $V$ channels initially remain constant before $M_P$ grows rapidly.
This is in agreement with the results obtained with staggered
fermions~\cite{Cheng:2010fe,Banerjee:2011yd}. Around $T_C$ the masses in $V$ and
$A$ channels become degenerate, in agreement with the restoration of the chiral
$SU_A(2)$ symmetry, and are about 10~\% smaller than the asymptotic value. In
contrast, the masses of $P$ and $S$ channels remain disparate, signaling
the persistent breaking of the $U_A(1)$ symmetry, and are in the region of 40 to
60~\% of the $2\pi T$ limit. Above $T_C$ all screening
masses approach the asymptotic limit from below, which might be a remnant of the
relatively small volume with $N_s/N_t=2$ (see~\cite{mueller_lattice}). Finite
size effects will be studied once the larger volumes become available. The
masses in $P$ and $S$ channels become degenerate at about $1.25\;T_C$ for scan
{\bf D1} (for scan {\bf C1} they are still non-degenerate at $T/T_C\approx1.17$)
which coincides with the result at $m_\pi=200$~MeV from $N_f=2+1$ domain
wall~\cite{bazavov:2012ja,Buchoff:2013nra} and staggered
fermion simulations~\cite{Cheng:2010fe,Banerjee:2011yd}.

\begin{figure}[t]
\vspace*{-5mm}
 \centering
\includegraphics[width=.65\textwidth]{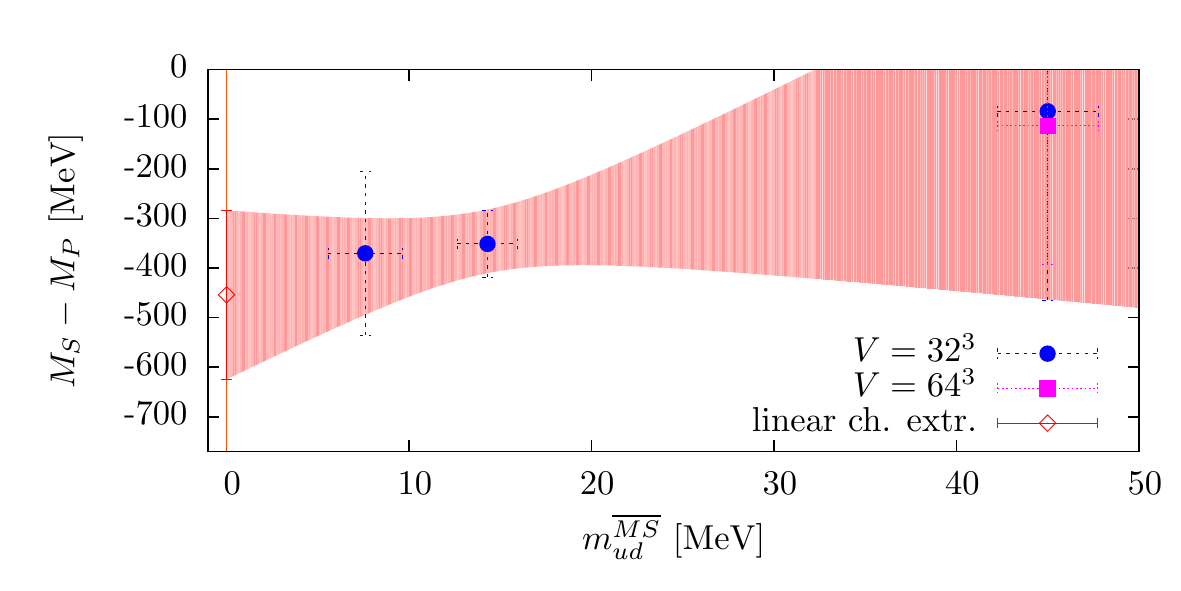}
\vspace*{-4mm}
 \caption{Chiral extrapolation of $M_S-M_P$ in MeV on the $32^3$ volume in the
thermal transition region using a linear function in $m_{ud}$. Also shown is the
result of scan {\bf B3}$_\kappa$ at a volume of $64^3$.}
 \label{fig:chext}
\end{figure}

Relevant for the order of the transition is the strength of the breaking of
$U_A(1)$ in the chiral limit which can be assessed by performing a chiral
extrapolation for some suitable measure. Here we use the average of the mass
difference in $S$ and $P$ channels in the transition region as a measure for the
strength of the breaking, which we then extrapolate to the chiral limit. The
results for scans {\bf B1}$_\kappa$, {\bf C1} and {\bf D1} are shown in
figure~\ref{fig:chext} together with a linear extrapolation in $m_{ud}$. As can
be seen from the plot, our current results strongly suggest a non-vanishing
difference at $m_{ud}=0$. However, to address the question of the order of the
transition one still has to clarify the criterion which determines a `strong'
breaking. It appears that a proper criterion should make contact with the
strength of the breaking at $T=0$, so that the breaking at $T_C$ can be
expressed in relation to this. A possible measure is provided by the mass of the
iso-vector scalar particle at $T=0$ in the chiral limit. A comparison with this
quantity will be done in future publications.

\section{Conclusions and perspectives}

In this proceedings article we have presented the current status of our
project to investigate the order of the transition in the chiral limit at
$N_f=2$. The available results for the transition temperatures at three
different pion masses between 200 and 500~MeV at $N_t=16$ allow for a first
test of critical scaling in the approach to the chiral limit. As expected, using
the scaling alone, with the current set of still relatively large pion masses,
is not sufficient to distinguish between the two scenarios. Scans at smaller
pion masses are in preparation and will hopefully improve the situation. As
another source of information we have extracted the difference of screening
masses in $S$ and $P$ channels which is a measure for the strength of the
breaking of the $U_A(1)$ symmetry. The first naive chiral extrapolation of this
quantity shows that the effect has the tendency to become stronger in the
approach to the chiral limit. However, simulations at smaller pion masses and at
larger volumes are needed to confirm this statement.

\noindent{\bf Acknowledgements:} The simulations were done on the WILSON and LC2
clusters at the University of Mainz, the FUCHS cluster at the University of
Frankfurt and on JUROPA and JUGENE
at FZ J\"ulich under project number HMZ21.

\end{document}